\documentclass[12pt,a4paper]{article}
\usepackage{amssymb,amsmath,graphicx,subcaption,hyperref,cite}
\usepackage{epstopdf,xcolor}
\usepackage[utf8]{inputenc}
\usepackage{color}
\usepackage[english]{babel}
\begin{document}

\begin{center}
 {\Large\bf Existence of two distinct time scales in the Fairen-Velarde model of bacterial respiration}
\end{center}
\vskip 1 cm
\begin{center} %
 Soumyadeep Kundu$^1$ and Muktish Acharyya$^{2,*}$
  
 \textit{Department of Physics, Presidency University,} 
  
 \textit{86/1 College Street, Kolkata-700073, India} 
 \vskip 0.2 cm
 {Email$^1$:sdpknu@gmail.com}
  
 {Email$^2$:muktish.physics@presiuniv.ac.in}
\end{center}
\vspace {1.0 cm}
\vskip 1.5 cm 

\noindent {\large\bf Abstract:} We study the bacterial respiration through the numerical solution of the
Fairen-Velarde coupled nonlinear differential equations. The instantaneous concentrations of the oxygen
and the nutrients are computed. The fixed point solution and the stable limit cycle are found in different 
parameter ranges as predicted by the linearized differential equations. In a specified range of parameters,
it is observed that the system spends some time near the stable limit cycle and eventually reaches the 
stable fixed point. This metastability has been investigated systematically. Interestingly, it is observed
that the system exhibits two distinctly different time scales in reaching the stable fixed points. The slow time scale of the
metastable lifetime near the stable limit cycle and a fast time scale (after leaving the zone of limit cycle) in rushing towards the
stable fixed point. The gross residence time, near the limit cycle (described by a slow time scale), can be reduced by varying the concentrations
of nutrients. This idea can be used to \textcolor{blue}{control} the harmful metastable lifespan of active bacteria.


\vskip 2cm

\noindent $^*$ Corresponding author
\newpage

\section{Introduction}

\textcolor{blue}{Microbiology, an important topic of modern research, deals with the behavior and respiration of bacteria, providing crucial insights for human survival in the contemporary world. Bacteria, among the earliest life forms on Earth, can thrive in diverse environments, including deep oceans\cite{Ocean}, hot springs\cite{Springs}, and even space\cite{Space}, where they can survive without oxygen. The evolution of oxygen in the atmosphere which was absent in the early atmosphere\cite{EarlyAtmosphere}, was demonstrated by Miller and Urey \cite{Miller}. Every living organism has to respire and generate energy to survive, so as Bacteria. Most organisms do their respiration in the presence of oxygen, but some of them can do it in its absence. The former process is called aerobic respiration and the latter process is called anaerobic respiration. The role of oxygen in aerobic respiration is as a final electron acceptor. In anaerobic respiration, the role is performed by sulphates, nitrites, etc\cite{Campbell}. In this paper, we shall talk about the genus \textit{Klebsiella},  part of the Enterobacteriaceae family \cite{Chart, Samanta}. We shall examine two species of \textit{Klebsiella}, \textit{K. aerogenes} and \textit{K. pneumonia}. This bacteria performs facultative anaerobic respiration\cite{Prescott} i.e. it can perform respiration in the absence of oxygen but will intake oxygen if it is present in the medium. However, the growth of this bacteria is extremely poor in strictly anaerobe conditions. Recent studies show a worrisome rise in antimicrobial resistance (AMR) in \textit{Klebsiella}\cite{Anti}, posing a threat to public health. Antibiotics, essential in combating bacterial infections, either slow bacterial growth (bacteriostatic) or directly kill bacteria (bactericidal) \cite{Kleb}. In a North Indian hospital from 2018 to 2022, a study focused on\textit{ K. pneumonia} found an increase in resistance to amoxicillin from 10\% in 2018 to none in 2022. Similarly, the rate of resistant \textit{K. pneumonia} rose from 7.5\% to 21.4\%, with extensively drug-resistant cases increasing significantly from 62.5\% to 71\% \cite{Ind}.}

\textcolor{blue}{We are taking the available experimental data of \textit{K. aerogenes} to understand the respiration process. The respiration rate of \textit{Klebsiella} is dependent on the concentration of nutrients and oxygen. Higher oxygen concentration inhibits the respiration procedure, as a result, we can observe a hysteresis in the oxygen tension in gas and liquid medium\cite{Degn}. We are examining the oxygen and glucose concentration in the medium. We shall check if there is excess glucose($\approx10$ mg/ml) or limited glucose($\approx2.5$ mg/ml) in the medium. Nitrogen is used as the growth limiting factor. In both cases, we are observing no oscillation in dissolved oxygen tension in excess oxygen($\approx~15$ mm Hg) or limited oxygen($<1$ mm Hg) state, though there is an oscillation between them.\cite{Pitt}. This oscillation will be shown as the formation of limit cycle in this paper.}

In the existing literature, we have not found any systematic study on the timescales of different phases of the bacteria.
\textcolor{blue}{This motivated us to explore the time scales in the transient phases (approaching fixed point) of bacteria.}
In this paper, we have taken the initiative to investigate the time scales of different phases of the bacteria. We have
considered celebrated Fairen-Velarde \cite{fairen} model of bacterial respiration and studied systematically. We have organised
the manuscript in the following format: The Fairen-Velarde model of bacterial respiration is introduced in the next section
(section-2). The mathematical analysis \textcolor{blue}{(in the linearized limit)} of the existence of stable limit cycle and stable fixed point is given in section-3. \textcolor{blue}{To incorporate the effects of nonlinearity,}
the Runge-Kutta-Fehlberg methodology of the numerical solution of the Fairen-Velarde model of a set of \textcolor{blue} {nonlinear coupled
differential equations} is described in section-4. The numerical
results are shown in section-5. The paper ends with a summary and concluding remarks in section-6.

\section{Fairen-Velarde Mathematical model of Bacterial respiration}

\noindent \textcolor{blue}{Theoretically, to study the lifecycles of the bacteria one should search for a suitable model. This model
must provide the time dependences of the amounts of the nutrient and the oxygen which are the basic ingredients for
the survival of the bacteria. The differential equations involving such two variables (concentrations of the oxygen and
the nutrients) may be a model for such dynamical evolution. Moreover, the interdependences of these two basic variables
should be present in such a model of the differential equation. In this context, the Fairen-Velarde (1979) model for the
respiration of bacteria is the most celebrated model.}

 This Fairen-Velarde model \cite{fairen} of bacterial respiration is expressed in the form of the two-variable coupled nonlinear deterministic differential equations given by:

\begin{align}
    \frac{dx}{dt} &= k_1 a-k_2 x -k_4 \frac{xy} {1+k_5 x^2} &=f(x,y) \nonumber \\
    \frac{dy}{dt} &= k_3 b -k_4 \frac{xy} {1+k_5 x^2} &=g(x,y).
    \label{fv}
\end{align}

\noindent Where, $x(t)$ is the concentration of oxygen and $y(t)$ is the concentration of nutrients in the system. $k_1$ represents the forward transport rate 
of oxygen from chamber to system, $k_2$ represents the backward transport rate of the same. $k_3$ is the supply rate of the nutrients from chamber to system. $k_4$ and $k_5$ determine the rate of oxygen and nutrient uptake by the bacteria. $a$ and $b$ are the concentrations of oxygen and nutrients in the chamber respectively.

\textcolor{blue}{Let us try to understand the behaviours of emergent dynamical phases of bacteria with analytical method. This will help us to know the responses of the bacteria to the parameters like supply of oxygen and nutrients to the system of
bacteria.}

\vskip 0.5cm
\section{Mathematical analysis for the fixed point and limit cycle in Fairen-Velarde model}

\noindent The Fairen-Velrade nonlinear system can be mathematically
analysed \cite{strogatz} in the linearized limit. 

\subsection{Non-dimensionalizing Fairen-Velarde differential equations}

\noindent \textcolor{blue}{As a usual practice, to study \cite{strogatz} the linearized version of nonlinear equations, we begin with the nondimensionalisation of 
the parameters and variables.}

Let us introduce the new set of variables as,
    \begin{equation*}
        \tau=k_2t; ~~~~\hat{x}=\frac{k_4}{k_2}x;~~~~ \hat{y}=\frac{k_4}{k_2}y; ~~~~\alpha=\frac{k_1 k_4}{k_2^2}a;~~~ \beta=\frac{k_3 k_4}{k_2^2}b;~~~~ \kappa=\frac{k_2^2}{k_4^2}k_5        
    \end{equation*}
   
   The above redefinition nondimesionalizes the form and 
   the original Fairen-Velarde coupled nonlinear differential equations become,
    \begin{align}
        \frac{d\hat{x}}{d\tau}&=\alpha-\hat{x}-\frac{\hat{x}\hat{y}}{1+\kappa \hat{x}^2}=f_1(\hat{x}, \hat{y}) \nonumber\\
        \frac{d\hat{y}}{d\tau}&=\beta-\frac{\hat{x}\hat{y}}{1+\kappa \hat{x}^2}=f_2(\hat{x}, \hat{y})
    \label{dimless-fv}
    \end{align}
   
    Now, the fixed point of eqn.(\ref{dimless-fv}) can be found by demanding;
    \begin{equation}
        f_1(\hat{x}, \hat{y})=0,~~~~f_2(\hat{x}, \hat{y})=0
    \end{equation}
    
The coordinates of the fixed point ($x^*, y^*$) becomes:

\begin{equation}
    x^*=\alpha-\beta, ~~~\\
    y^*=\frac{\beta(1+\textcolor{blue}{\kappa}(\alpha-\beta)^2)}{\alpha-\beta}
    \label{fixedpoint}
\end{equation}

\textcolor{blue} {What is the physical significance of this fixed point? The fixed point provides the fixed values of the
oxygen concentration ($x^*$) and the concentration of nutrients ($y^*$). That means, that for a specified set of values of the
supply rate of oxygen and nutrient concentrations, the system eventually reaches a state of fixed point having fixed values
of oxygen and nutrient concentrations in the system. In this phase, the life cycle of the bacteria is arrested. Further, the
growth of bacteria is not possible and hence the desired phase for  mankind.} 

\subsection{Linearization of nonlinear equations}

\noindent \textcolor{blue}{In some cases, the linearized version of nonlinear differential equations can capture the basic outcomes. Keeping this in mind, let us linearize the Fairen-Velarde (eqns-\ref{dimless-fv}) coupled nonlinear differential equations with the compact vector notations
\cite{strogatz},}
 
\begin{equation}
    f(x,y)=\begin{pmatrix}
        f_1(x, y)\\
        f_2(x, y)
    \end{pmatrix}
\end{equation}

\noindent The Taylor's series expansion around the fixed point ($x^*,y^*$) leads to\cite{strogatz}:

\begin{equation}
    f(x,y)=f(\hat{x}^*,\hat{y}^*)+J(\hat{x}^*,\hat{y}^*)\begin{vmatrix}
        \hat{x}-\hat{x}^*\\
        \hat{y}-\hat{y}^*
    \end{vmatrix}+{\mathcal O} \left(\begin{Vmatrix}
        \hat{x}-\hat{x}^*\\
        \hat{y}-\hat{y}^*
    \end{Vmatrix}\right)
\end{equation}
\begin{equation}
    \frac{d\vec{u}}{d\tau}=J(\hat{x}^*,\hat{y}^*)\vec{u}; {\rm ~~where~~} {\vec u} = \begin{vmatrix}
        \hat{x}-\hat{x}^*\\
        \hat{y}-\hat{y}^*
    \end{vmatrix}
\end{equation}

\noindent Now, evaluating the Jacobian ($J$) at the fixed point ($x^*,y^*$),

\begin{align}
    J & = \begin{vmatrix}
        \frac{\partial f_1}{\partial \hat{x}} & \frac{\partial f_1}{\partial \hat{y}}\\
        \frac{\partial f_2}{\partial \hat{x}} & \frac{\partial f_2}{\partial \hat{y}}
    \end{vmatrix}\\
   &  = \begin{vmatrix}
        -1-\big(\frac{1-\kappa\hat{x}^2}{(1+\kappa\hat{x}^2)^2}\big)\hat{y}  & -\frac{\hat{x}}{1+\kappa\hat{x}^2}\\
        -\big(\frac{1-\kappa\hat{x}^2}{(1+\kappa\hat{x}^2)^2}\big)\hat{y} & -\frac{\hat{x}}{1+\kappa\hat{x}^2}
    \end{vmatrix};\\
    J(\hat{x}^*,\hat{y}^*)&=\begin{vmatrix}
        \frac{\beta(1-\kappa(\alpha-\beta)^2)}{(1+\kappa(\alpha-\beta)^2)(\beta-\alpha)}-1 & \frac{\beta-\alpha}{1+\kappa(\alpha-\beta)^2}\\
        \frac{\beta(1-\kappa(\alpha-\beta)^2)}{(1+\kappa(\alpha-\beta)^2)(\beta-\alpha)} & \frac{\beta-\alpha}{1+\kappa(\alpha-\beta)^2}
    \end{vmatrix}\\
\end{align}
Demanding the solution for stable fixed point,
\begin{align}
    Tr(J)&=J_{11}+J_{22}<0 \\
    det(J)&=J_{11}J_{22}-J_{12}J_{21}>0
\end{align}

\noindent If $\lambda$ represents the eigenvalues ($\lambda_1, \lambda_2$), then,
\begin{align}
    \lambda^2-Tr(J)\lambda+det(J)&=0\\
    \Rightarrow \lambda_{1,2}&=\frac{1}{2}(Tr(J)\pm \sqrt{(Tr(J))^2-4det(J)})
\end{align}

\noindent If both eigenvalues are real, $Tr(J)^2-4det(J)\geq 0$.
\begin{align}
    \lambda_1&=\frac{1}{2}(Tr(J)+\sqrt{(Tr(J))^2-4det(J)})\\
    \lambda_2&=\frac{1}{2}(Tr(J)-\sqrt{(Tr(J))^2-4det(J)})
\end{align}
Now, $Tr(J)<0$, $det(J)>0$. And, the modulus of the square-root part is less than the modulus of $Tr(J)$. So we shall get $\lambda_1<0; \lambda_2<0$.

\noindent If both eigenvalues are imaginary, $Tr(J)^2-4det(J)\le 0$.
\begin{align}
    \lambda_1&=\frac{1}{2}(Tr(J)+i\sqrt{4det(J)-(Tr(J))^2}\\
    \lambda_2&=\frac{1}{2}(Tr(J)-i\sqrt{4det(J)-(Tr(J))^2}
\end{align}
For stability (provide the shrinking spiral), the real part should be negative.
Now,
\begin{align}
    det(J)=\begin{vmatrix}
        J_{11} & J_{12}\\
        J_{21} & J_{22}
    \end{vmatrix}=\begin{vmatrix}
        J_{21}-1 & J_{22}\\
        J_{21} & J_{22}
    \end{vmatrix}=-J_{22}=\frac{\alpha-\beta}{1+\kappa(\alpha-\beta)^2}>0.
\end{align}
As the denominator is positive definite, from numerator we can get, $\alpha > \beta$.
Putting this condition in trace equation,
\begin{align}
    \kappa < \frac{\beta+(\alpha-\beta)+(\alpha-\beta)^2}{(\alpha-\beta)^2(2\beta-\alpha)}
\end{align}
$\kappa$ is positive definite, as it depends on rate constants only. As the numerator is positive definite, from the
denominator, we get $\alpha<2\beta$.

\subsection{Ranges of parameters}
All the parameters $\kappa, \alpha, \beta$ are positive definite because (since every term is dependent upon rate constants). So our conditions become,
\begin{align}
    0&<\beta<\alpha<2\beta\\
    0&<\kappa < \frac{\beta+(\alpha-\beta)+(\alpha-\beta)^2}{(\alpha-\beta)^2(2\beta-\alpha)}
    \label{final-parameters}
\end{align}
The above relations should be maintained to have a stable fixed point. The range of $\kappa$ outside the range given above will give the limit cycle.
We shall check the parameters in different ranges to check the stability(or limit cycle). We shall simulate the nonlinear equation (eqn.(\ref{dimless-fv})) by 6-th order Runge-Kutta-Fehlberg  method using the parameters (eqn.(\ref{final-parameters})) . \textcolor{blue}{The values of $\kappa$ outside the range shown above would provide the limit cycle. The limit cycle arises
due to steady oscillations of the concentrations of oxygen and nutrients with a phase difference. This is the proactive phase
of the bacteria. Bacteria generally prefer this phase to maintain their growing lifecycles. This phase is obviously not
desirable to mankind. The basic target of the research on  bacteria should be to make such life cycles a transient phase
which eventually leads to a phase of fixed point(favourable for mankind).}

\section{Numerical Solution of Fairen-Velarde coupled nonlinear differential equations}

\noindent \textcolor{blue}{The linearized solution can provide a reasonable solution. From the linearized analytical
solutions (mentioned above) we have an idea of the domain of parameter values which provide the stable limit cycle and
the stable fixed point. However, through  linearization, we have not captured the behaviour arising from nonlinearity.
To get the actual values of the parameters responsible for stable limit cycle and stable fixed point one must consider
the actual nonlinear coupled differential equations. These equations cannot be solved analytically. One needs the numerical
method of solving such nonlinear coupled differential equation.}

\vskip 0.5 cm

We have used the \textit{6th order Runge-Kutta-Fehlberg} technique\cite{wheatley} to solve the nonlinear coupled differential
equations (eqns.-\ref{dimless-fv}). We have taken the time interval $d\tau = 0.001$ so that the error involved in 6th order RKF method is quite small.

\vskip 1cm

\section{Numerical results}

\begin{figure*}[htpb]
\centering
\includegraphics[width=0.4\columnwidth]{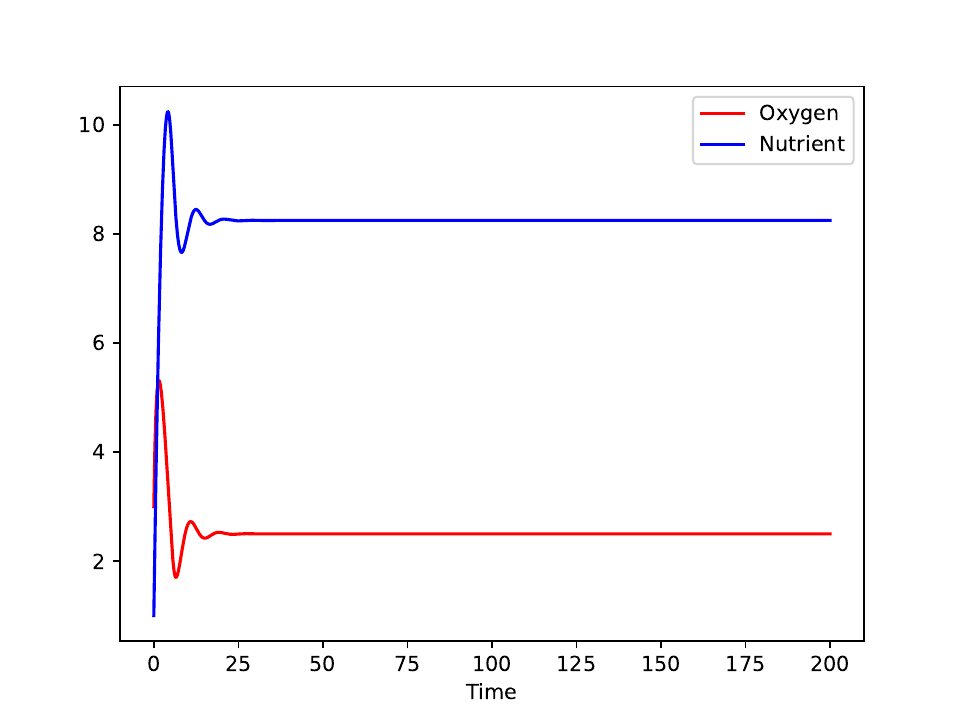}
(a)
\includegraphics[width=0.4\columnwidth]{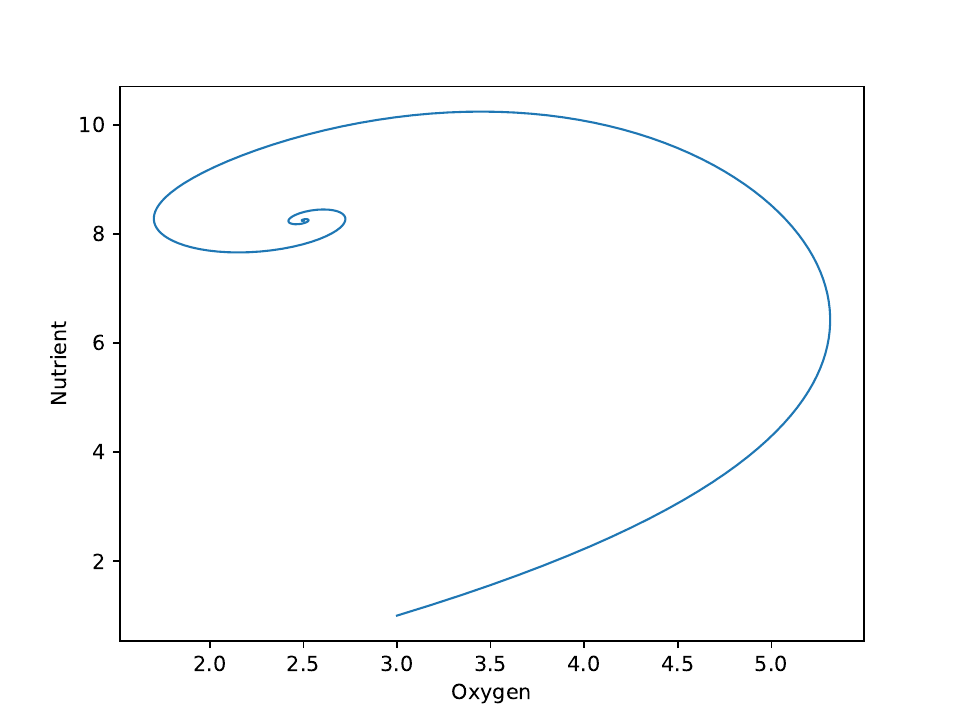}
(b)
\caption{The temporal evolution of the concentrations of the oxygen and nutrients. (a) the concentrations of oxygen and the nutrients are shown as functions of time and (b) the time eliminated form shows the
eventual achievement of the fixed point. Here, $\alpha=7.5$, $\beta=5.0$ and $\kappa=0.5$.}
\label{fp-simple}
\end{figure*}

We have solved the eqn.(\ref{dimless-fv}) by 6-th order Runge-Kutta-Fehlberg method\textcolor{blue}{\cite{wheatley}} for the fixed set of values of the parameters and the results are shown in Fig.\ref{fp-simple}. In these values of the  
parameters ($\alpha=7.5$, $\beta=5.0$ and $\kappa=0.5$, obeyed by eqn.(\ref{final-parameters}) as the condition of having stable fixed point), we
have shown that the existence of the stable fixed point. \textcolor{blue}{Since the lifecycle of the bacteria is arrested, the phase of stable fixed point is the desired phase for mankind.}

\vskip 1cm

\begin{figure*}[htpb]
\centering
\includegraphics[width=0.4\columnwidth]{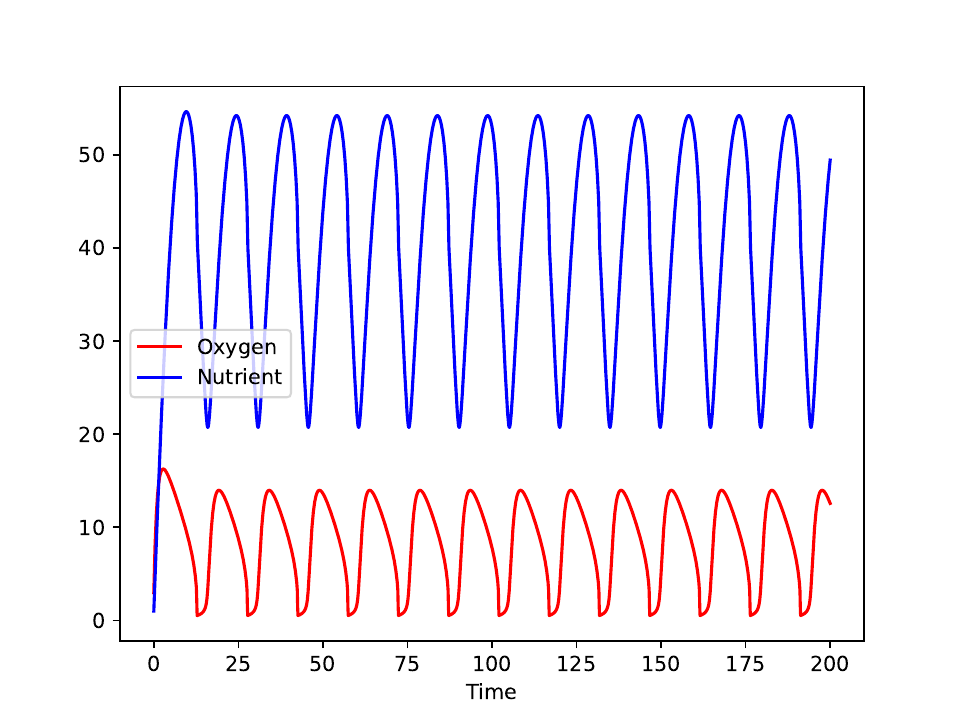}
(a)
\includegraphics[width=0.4\columnwidth]{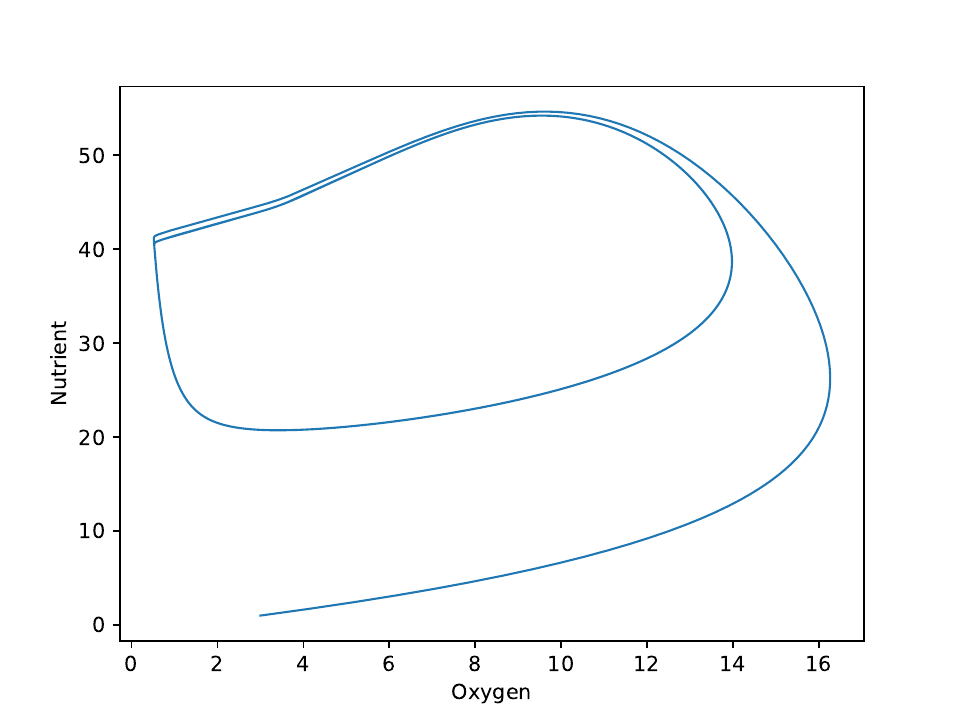}
(b)
\caption{The temporal evolution of the concentrations of the oxygen and nutrients. (a) the concentrations of oxygen and the nutrients are shown as functions of time and (b) the time eliminated form shows the
eventual achievement of the stable limit cycle. Here, $\alpha=19.4$, $\beta=11.0$ and $\kappa=0.5$.}
\label{lc-simple}
\end{figure*}

On the other hand, the stable limit cycle may exist in different sets of values of the parameters ($\alpha=19.4$, $\beta=11.0$ and $\kappa=0.5$, which do not obey the condition given in eqn.(\ref{final-parameters})), where stable unattenuated
oscillations of the concentrations of oxygen and nutrients are observed. 
The oscillation of instantaneous concentration of the oxygen 
($x(t)$) in the system, is due to the availability of the nutrient
($y(t)$). If the instantaneous concentration of nutrient ($y(t)$) 
is increasing then it means the consumption by the bacteria is decreasing
due to the death of bacteria. This shows the increase in oxygen concentration
also, having a phase lag. The time eliminated for of the oscillations of
oxygen and nutrients will provide the limit cycle. Since the number of bacteria decreased, there was an adequate amount of oxygen in the system, so the bacteria started to respire and reproduce, and as a result, consumption of oxygen and  nutrients increased which is shown by the decrease of the concentrations in the system.
This existence of limit cycle ensures the active
phases of the bacteria. These results are shown in Fig.\ref{lc-simple}. \textcolor{blue}{ This stable limit cycle 
corresponds to the stable and active lifecycles of the bacteria. This phase is harmful to  mankind and usually not
desirable. The research on bacteria must provide the values of the parameters that may lead to the stable limit cycle.
The people should be aware of such conditions represented by the values of the parameters. We always have to avoid
such domain of parameter values.}

\begin{figure*}[htpb]
\centering
\includegraphics[width=0.4\columnwidth]{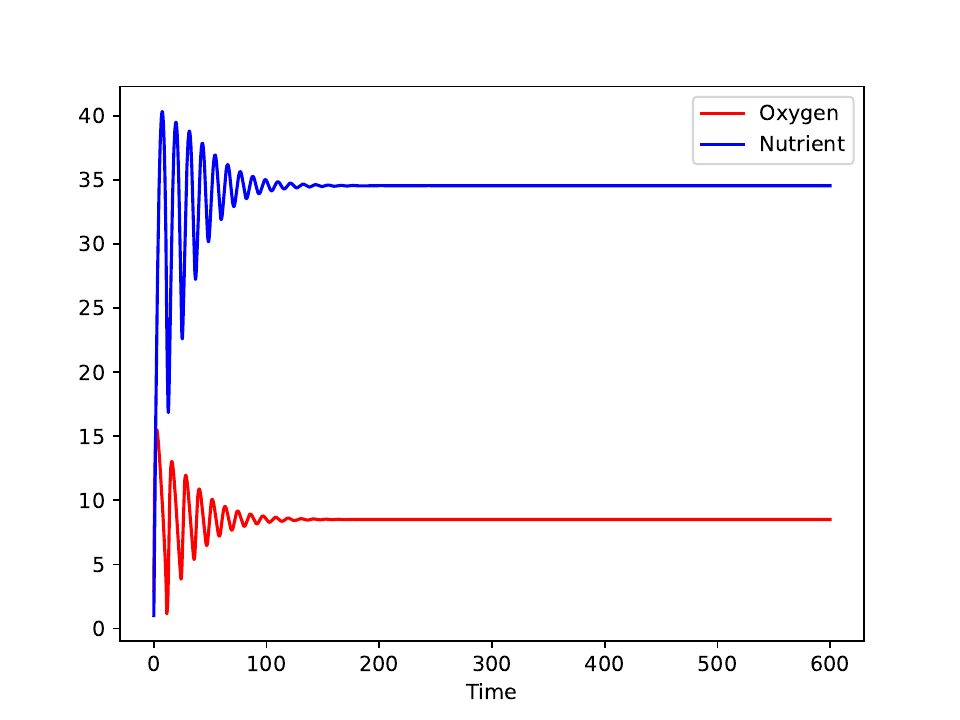}
(a)
\includegraphics[width=0.4\columnwidth]{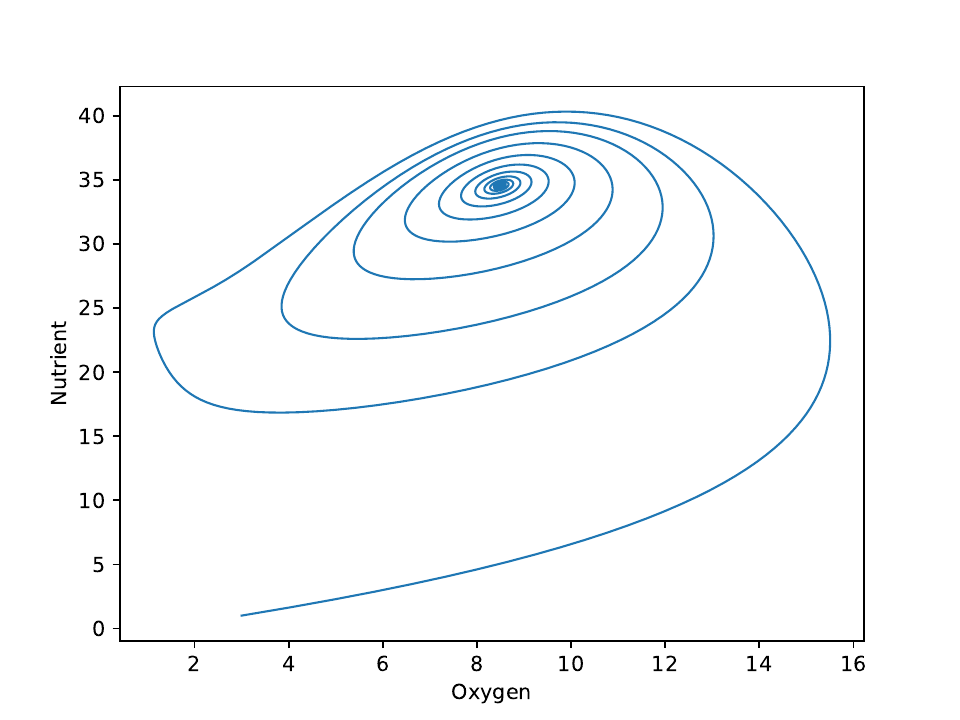}
(b)
\caption{(a) The transient oscillations of the concentrations of the oxygen and nutrients. 
(b) Time eliminated dynamics of the concentrations of the oxygen and nutrients. The system eventually 
reaches the stable fixed point after several orbiting around it. Here, $\alpha=19.4$, $\beta=10.9$ and $\kappa=0.359$.}
\label{meta}
\end{figure*}

We have identified a specified range of parameter values where the system shows the metastable behaviour before reaching the
stable fixed point. This metastability is basically a temporal residence near the limit cycle. The system spends a considerable
amount of time near the limit cycle, then leaves this metastable state and eventually reaches the stable fixed point.
 \textcolor{blue}{This metastable phase is better than the phase of the stable limit cycle. We know that the decay of such a
 metastable phase would eventually lead to the phase of stable fixed point. The phase of stable fixed point is desirable
 phase for mankind.} In the
language of bacteria, the fixed point does not show any time dependences of the concentrations of oxygen as well as the 
concentration of nutrients. As a consequence, the activity of the bacteria gets lost. One such metastable behaviour is
 shown in Fig.\ref{meta} for the parameter values $\alpha=19.4$, $\beta=10.9$ and $\kappa=0.359$. It may be noted here, from the condition
 (eqn.(\ref{final-parameters})) of parameters $\kappa$ should be less than 
 0.4904 to give a fixed point. However, we are getting the limit cycle even
 for $\kappa=0.359$ (which apparently violates the condition). The reason behind it is the simplification due to the linearization of Fairen-Velarde nonlinear coupled differential equations.
 \textcolor{blue}{Now, the basic question is how long the metastability persists. Or how long we have to wait in a
 phase of unstable limit cycle? For this purpose, one has to know about the time scale of such metastable phase. If,
 the research can provide the idea of such lifetime of metastable phase, the time to get cured of infectious disease
 can be predicted.}
 
\begin{figure*}[htpb]
\centering
\includegraphics[width=0.4\columnwidth]{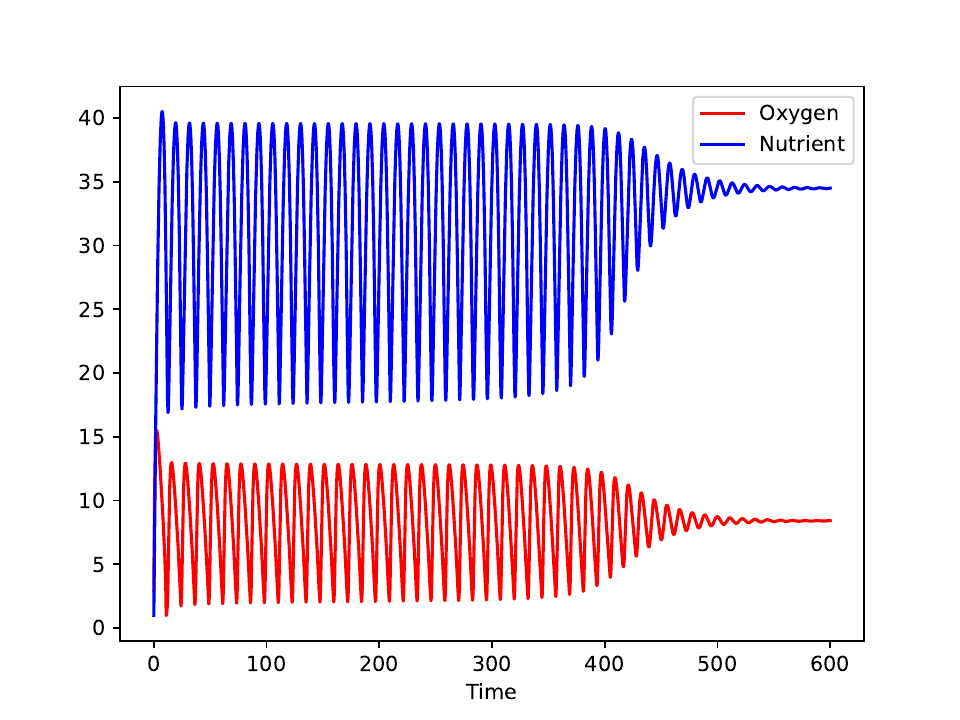}
(a)
\includegraphics[width=0.4\columnwidth]{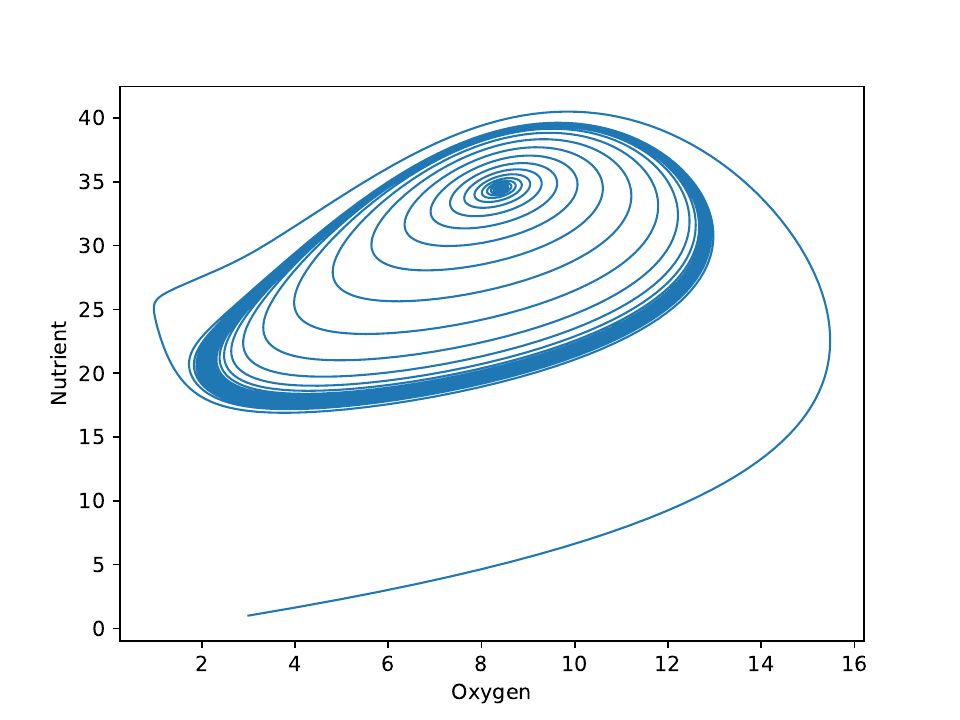}
(b)
\caption{(a) The transient oscillations of the concentrations of the oxygen and nutrients. 
(b) Time eliminated dynamics of the concentrations of the oxygen and nutrients. The system eventually 
reaches the stable fixed point after several orbiting around it. Here, $\alpha=19.4$, $\beta=10.997$ and $\kappa=0.359$.}
\label{long-meta}
\end{figure*}

Keeping the fixed values of the parameter $\alpha=19.4$ and $\kappa=0.359$, the longer metastability is observed for a slightly
larger value of $\beta$ (=10.997). It may be noted here, the metastable lifetime can be externally \textcolor{blue}{controlled} by the renormalized concentration of the nutrients ($\beta$). One such result are depicted in Fig.\ref{long-meta}. The longer
metastability (oscillatory behaviour) is clear as compared to that shown in Fig.\ref{meta}. The thick region near the limit
cycle (in Fig.\ref{long-meta}(b)) also indicates the longer metastability here. The system stays over a significant period of time near the limit cycle before rushing towards the stable fixed point.

\begin{figure*}[htpb]
\centering
\includegraphics[width=0.4\columnwidth]{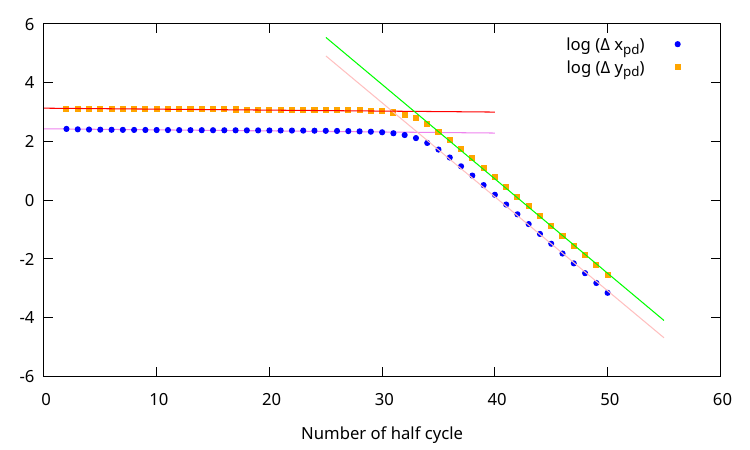}
(a)
\includegraphics[width=0.4\columnwidth]{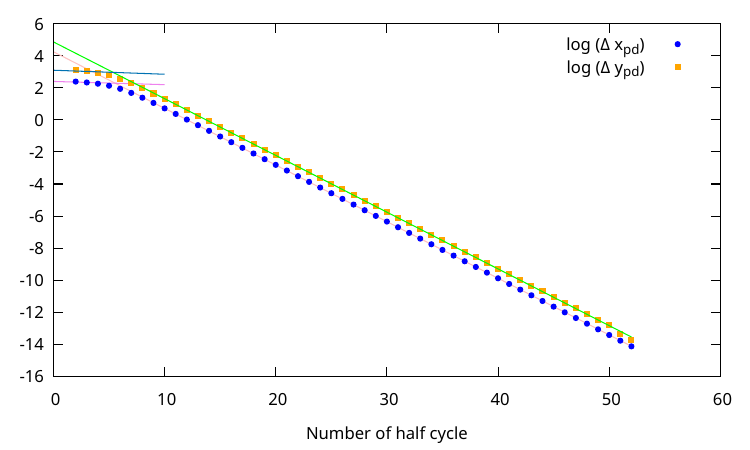}
(b)
\caption{The logarithm of the absolute differences ($\Delta x_{pd}$ and $\Delta y_{pd}$)  of
the values of consecutive peak and dip are plotted against the number of half period of oscillations. (a) high metastbale
lifetime for $\beta=10.997$ and (b) low metastable lifetime for $\beta=10.980$. The straight line in the semilogarithmic
plot indicates the exponential decay. The continuous line represents the 
best fit (exponentially decaying).}
\label{timescales}
\end{figure*}

{\it Can one get any idea about the scale of such a metastable lifetime near the limit cycle?} We have tried to get such a
timescale from the age-old knowledge of damped harmonic oscillators. The amplitude of the underdamped harmonic oscillator
decays exponentially. The inverse of that decay constant provides the scale of time. We have employed the same notion
here and found the scale of metastable lifetime near the limit cycle. From the oscillation of the concentrations of the
oxygen and the nutrients, the logarithm of the absolute difference ($\Delta x_{pd}$ and $\Delta y_{pd}$) of the values of consecutive peak and dip of $x(t)$ (or $y(t)$) is studied as a function of the number of half periods of oscillation($n_p$). 
The results are shown in Fig.\ref{timescales} for two different values of $\beta$ keeping $\alpha=19.4$ and $\kappa=0.359$.
The straight line in semi-logarithmic plot indicates the exponential decay ($e^{-t/T}$ type), which in turn provides the scale of time ($T$). Initially,
the system resides in the metastable states (near the limit cycle). This corresponds to the lower value of slope (high $T$) of the straight line in semilogarithmic plot. On the other hand, the system leaves the metastable state and rushes towards the stable fixed point, with faster rate (indicated by the
higher value of the slope (low $T$) of the straight line in the semilogarithmic plot. The scale of exponential decay of the metastable lifetime
($T_{meta}$) and the scale of exponentially achieving the fixed point ($T_{fixed}$) are different. In our simulation, we have obtained
$T_{fixed}~ < ~ T_{meta}$. Moreover, the gross span of the residence time at the metastable state (span of lower slope of the straight line in the plot) can be lowered by varying $\beta$ (the renormalised concentration of the nutrients). Consequently,
the faster achievement of the fixed point can be \textcolor{blue}{controlled} to destroy the active cycles (limit cycles) of the bacteria.

\textcolor{blue}{The emergent two distinct scales of metastable lifetimes have a far
reaching consequence. Mainly, the larger timescale, close to the limit
cycle keeps the bacteria in active phase. The main target is to reduce this
time scale to accelerate the metastable life towards the fixed point. 
The medication or any other control mechanism would be called successful
if it can reduce the active period of bacterial life cycle. The main 
findings of this artical is to find the parameter which can reduce this
longer timescale. The parameter $\beta$ plays the crucial role here. As
$\beta$ decreases, this longer timescale reduces (Fig-\ref{timescales}).}

\newpage

\section{Summary and concluding remarks}

\noindent Fighting against various harmful bacteria is a major challenge to modern civilization. The life cycle,
respiratory behaviour etc. should be targetted to fight against it. Apart from the biologically \textcolor{blue}{controlled}
experimental survey, one should know the cyclic behaviour of bacterial lifecycle. Mostly, the theoretical
understanding and quantitative analysis via the mathematical model demand serious scientific attention.

In this regard, we have thoroughly investigated the Fairen-Velarde (FV) mathematical model dedicated to
bacterial respiration. The model is described by two coupled nonlinear differential equations for the
concentrations of oxygen(required for aerobic respiration) and nutrients (required as food). We have
studied the FV nonlinear differential equations in the linearized (around the stable fixed point) case
analytically to explore the existence of the limit cycles. We have also solved the FV nonlinear coupled
differential equations numerically by Runge-Kutta-Fehlberg method to get such a stable limit cycle and the
stable fixed point. The stable limit cycle is a manifestation of the steady oscillation of the concentration
of the oxygen as well as the concentration of the nutrient. In the stable limit cycle the bacteria are in 
active phase. This active phase of bacterial lifecycle is harmful to the human race. The limit cycle is
observed in a range of values of the parameters. On the other hand, in some range of parameters, the limit
cycle becomes unstable and a stable fixed point may be the ultimate fate of the bacterial system. In the
fixed point, the concentrations of oxygen and that of the nutrient become time independent. In this
phase, the bacteria are not active. This phase brings good news for us.

In our study, our major findings are the existence of two different time scales. The  metastable state,
near the limit cycle, decays exponentially ($e^{-(t/T_{meta})}$). After leaving the metastable state, the
bacterial system rushes towards the stable fixed point, also exponentially ($e^{-(t/T_{fixed})}$), but with a faster rate. We
have found $T_{fixed} < T_{meta}$. Moreover, rather more interestingly, the total residence time in the
metastable phase can be varied (lowered) by varying $\beta$. 

The reduction of the total residence time near metastable state by varying $\beta$ reduces the active (harmful)
period of the bacteria. This can be used in biomedical controlling of bacteriology \textcolor{blue}{towards the better
and healthy situations of mankind infected by bacteria.}
It would be interesting to see the effects experimentally and to find the
real time scales.

\textcolor{blue}{Last but not the least, it is worth mentioning here, a more plausible model
would have some stochasticity included.
For example, adding a noisy perturbation (with specified and reasonable statistical distribution) to the equations would probably
affect the lifetime around the limit cycle.
In this context, it is worth exploring the effect of noise (arising from the environment) on the transition
time between the limit cycle and fixed point.
Of course this would also require a different approach in integrating the
equations with more efficient computational drive. The effects of the externally induced noise may also be studied to make the metastable
phase shorter lived.}
\newpage

\noindent {\bf Acknowledgements:}

\noindent \textcolor{blue}{We would like to thank the anonymous reviewer of this paper for sincere review with many important
suggestions. We thank Swagata Chattopadhyay, Calcutta University, for a careful reading of the manuscript.} SDK thankfully acknowledges the library facility of Presidency University, Kolkata. 

\vskip 0.5cm

\noindent {\bf Authors' contribution:}

\vskip 0.2cm

\noindent Soumyadeep Kundu has developed the code for numerical calculations, collected data, prepared figures and written the manuscript. Muktish Acharyya has conceptualized the problem, analysed the results and writen the manuscript. 

\vskip 0.3cm

\noindent {\bf Data availability statement:}
The data will be available on reasonable request to Soumyadeep Kundu.

\vskip 0.3cm

\noindent {\bf Conflict of interest statement:} We declare that this manuscript is free from any conflict of
interest. The authors have no financial or proprietary interests in any material discussed in this article.

\vskip 0.3cm
\noindent {\bf Funding statement:} No funding was received particularly to support this work.

\vskip 1cm

\newpage

\end{document}